\documentclass[aps,prl,showpacs,nofootinbib,twocolumn,float,superscriptaddress]{revtex4-1}
\usepackage{amsmath,bm,epsfig}
\newcommand{\B}[1]{{\bm{#1}}}
\newcommand{\C}[1]{{\mathcal{#1}}}    

\begin{document}
\title{Turbulence in non-integer dimensions by fractal Fourier
  decimation} \author{Uriel Frisch}
\affiliation{UNS,~CNRS,~OCA,~Laboratoire~Cassiop\'ee,~B.P.~4229,~06304~Nice~Cedex~4,~France}
\author{Anna Pomyalov} \affiliation{Department of Chemical Physics,
  The Weizmann Institute of Science, Rehovot 76100, Israel}
\author{Itamar Procaccia} \affiliation{Department of Chemical Physics,
  The Weizmann Institute of Science, Rehovot 76100, Israel}
\author{Samriddhi Sankar Ray}
\affiliation{UNS,~CNRS,~OCA,~Laboratoire~Cassiop\'ee,~B.P.~4229,~06304~Nice~Cedex~4,~France}
\altaffiliation{Also at Centre for Condensed Matter Theory, Department
  of Physics, Indian Institute of Science, Bangalore, India}

\begin{abstract}
Fractal decimation reduces the effective dimensionality of a flow by keeping
only a (randomly chosen) set of Fourier modes whose number in a ball of radius
$k$ is proportional to $k^D$ for large $k$. At the critical dimension $D=4/3$
there is an equilibrium Gibbs state with a $k^{-5/3}$ spectrum, as in
[V.~L'vov {\it et al.},  Phys. Rev. Lett. {\bf 89}, 064501 (2002)]. Spectral
simulations of fractally decimated two-dimensional turbulence show that the
inverse cascade persists below $D=2$ with a rapidly rising Kolmogorov
constant, likely to diverge as $(D-4/3)^{-2/3}$.
\end{abstract}
\date{\today} 
\maketitle 
In theoretical physics a number of
interesting results have been obtained by extending the dimension $d$
of space from directly relevant values such as 1, 2, 3 to non-integer
values. Dimensional regularization in field theory
\cite{thooft-veltman} and the $4-\epsilon$ expansion in critical
phenomena \cite{wilson-fisher} are well-known instances. In such
approaches, one usually expands the solution in terms of Feynman
diagrams, each of which can be analytically continued to real or
complex values of $d$. The
same kind of extension can be carried out for homogeneous isotropic
turbulence but a severe difficulty appears then for $d<2$:
the energy spectrum $E(k)$ can become negative in some band of
wavenumbers $k$, so that this kind of extension lacks probabilistic
reallizability \cite{fournier-frisch}.  Nevertheless, in
Ref.~\cite{LPP}, henceforth cited as LPP, it is argued that, should
there exist an alternative realizable way of doing the extension below
dimension two in which the nonlinearity conserves energy and
enstrophy, then an interesting phenomenon -- to which we shall come
back -- should happen in dimension 4/3.

For diffusion and phase transitions there is a very different way of switching
to non-integer dimensions, namely to reformulate the problem on a fractal of
dimension $D$ (here a capital $D$ will always be a fractal dimension)
\cite{diffusion-phase-transitions-fractals}. Are we able to do this for
\textit{hydrodynamics}? Implementing mass and  momentum conservation on a
fractal is quite a challenge.\footnote{
Lattice Boltzmann models may be amenable to fractal
  decimation \cite{hchen-private-communication}.} We discovered a new 
way of fractal decimation in Fourier space, appropriate for hydrodynamics.
Since, here, we are primarily interested in dimensions less than two, we shall
do our decimation starting from the standard $d=2$ case.

 The forced incompressible Navier-Stokes equations for the velocity field
can be written in abstract notation as
\begin{eqnarray}
\partial_t \B u& =& B(\B u,\B u) +\B f +\Lambda \B u\ , \label{NS}\\
B(\B u,\B u) &=& - \B u\cdot \B \nabla \B u +\B \nabla p\ , \quad \Lambda =\nu
\nabla^2 \ ,
\end{eqnarray}
where $\B u$ stands for the velocity field $\B u(x_1,x_2,t)$,
$\B f$ for $\B f(x_1,x_2,t)$, $p$ is the pressure and $\nu$ the viscosity. The velocity $\B u$ is taken in the space of
divergence-less velocity fields which are $2\pi$ periodic in $x_1$ and
$x_2$, such that $\B u(t=0)=\B u_0$.  Now, we define a Fourier
decimation operator $P_D$ on this space of velocity fields:
\begin{equation}
{\rm If}\,\,\,\,\B u =\sum_{\B k\in \C Z^2} e^{i\B k\cdot\B x} \hat
{\B u}_{\B k}, \,\,\,\,{\rm then}\,\,\, \,P_D \B u =\sum_{\B k\in \C
  Z^2} e^{i\B k\cdot\B x}\theta_{\B k} \hat {\B u}_{\B k} \ .
\end{equation}
Here $\theta_{\B k}$ are random numbers such that
\begin{equation}
\theta_{\B k} = \begin{cases}1 ~ \text{with probability}~h_k\\0
  ~\text{with probability}~1-h_k \ , \quad k\equiv |\B k| \
  .\end{cases}
\end{equation}
To obtain $D$-dimensional  dynamics  we choose
\begin{equation}
h_k = C (k/k_0)^{D-2} \ , \quad 0<D\le2, \quad 0< C\le 1 \ ,
\end{equation}
where $k_0$ is a reference wavenumber; here $C=k_0 =1$.
All the $\theta_{\B k}$ are chosen independently, except that
$\theta_{\B k}=\theta_{-\B k}$ to preserve Hermitian symmetry.
Our fractal decimation procedure removes at random --- but in a time-frozen (quenched)
way --- many modes from the $\B k$ lattice,  leaving on average 
$N(k)\propto k^D$ active modes in a disk of radius $k$. The randomness in the choice of the decimation will be
called the disorder.\footnote{A more drastic and non-random decimation is
the reduced wave-vector set approximation (REWA; see
Ref.~\cite{grossmann-lohse-reeh96} and references therein), in which the
number of active mode grows as $\ln k$, so that from our point of view it
has dimension $D=0$.} 

Observe that $P_D$ is a projector, that it commutes with the viscous
diffusion operator $\Lambda$ and that it is self-adjoint for the energy
($L^2$) norm, defined as usual as $||{\B u}||^2 \equiv (1/(2\pi)^2)
\int |{\B u}({\B x})|^2 \, d^2x$, where the integral is over a
$2\pi\times 2\pi$ periodicity square.  The conservation of energy (by
the nonlinear term) for sufficiently smooth solutions of the
Navier--Stokes equation can be expressed as $({\B u},\, B(\B u,\B u))
=0$ where $({\B u},\, {\B w}) \equiv (1/(2\pi)^2) \int {\B u}({\B
  x})\cdot {\B w}({\B x}) \, d^2x$ is the $L^2$ scalar product.

The \textit{decimated Navier--Stokes equation}, written  for an incompressible 
field $\B v$, takes the following form:
\begin{equation}
\partial_t \B v = P_D B(\B v,\B v) +P_D \B f +P_D \Lambda \B v\ . \label{NSD}
\end{equation} 
The initial condition is $\B v_0 \equiv \B v(t=0)=P_D \B u_0$. Thus
at any later time $P_D \B v =\B v$. Energy is again conserved; indeed  
$({\B v},\, P_DB(\B v,\B v)) =0$, as is seen by moving the self-adjoint
operator $P_D$ to the left hand side of the scalar product and using
$P_D{\B v} = {\B v}$. Enstrophy conservation for the decimated problem is
proved in a smilar way by working with the vorticity. 

 If, in addition to decimation, we apply a Galerkin truncation which
kills all the modes having wavenumbers beyond a threshold $K_G$, the surviving
modes constitute a dynamical system having a finite number of degrees of freedom. Such
truncated systems with no forcing and no viscosity have been studied by Lee, Kraichnan and others \cite{abseq}. Using suitable variables related
to the real and imaginary parts of the active modes, the dynamical
equations may be written as
$\dot y_\alpha =\sum_{\beta \gamma} A_{\alpha\beta\gamma} y_\beta y_\gamma$.

For the purely Galerkin-truncated (not decimated) case it is well
known that the above dynamical system satisfies a Liouville theorem
$\sum_\alpha \partial \dot y_\alpha/\partial y_\alpha =0$ and thus
conserves volume in phase space. This in turn implies the existence of
(statistically) invariant Gibbs states for which the probability is a
Gaussian, proportional to $e ^{-(\alpha E +\beta \Omega)}$ where $E =
\sum_{\B k} |\hat {\B u}_{\B k}|^2$ is the energy and $\Omega =
\sum_{\B k} k^2|\hat {\B u}_{\B k}|^2$ is the enstrophy.  Such
\textit{Gibbs states}, called by Kraichnan \textit{absolute
  equilibria}, play an important role in his theory of the
two-dimensional (2D) inverse energy cascade \cite{kraichnan67}. If we
now combine inviscid, unforced Galerkin truncation and decimation, it
is easily checked that the Liouville theorem still holds,
provided the decimation preserves Hermitean symmetry. 
For such Gibbs states, and
any active mode ($\theta_{\B k} =1$), one easily checks that the mean
square energy $\langle |{\B u}_{\B k}|^2\rangle = C'/(\alpha+\beta
k^2)$, where $C'>0$ does not depend on ${\B k}$.  The corresponding
energy spectrum is the mean energy $E(k)$ of modes having a wavenumber
between $k$ and $k+1$. Up to fluctuations of the disorder, the number
of active modes in such a shell is $O(k^{D-1})$. Thus,
\begin{equation}
E(k) = \frac{k^{D-1}}{\alpha+\beta k^2}; \quad \beta > 0, \quad
\alpha > -\beta,
\label{Ddimabseq}
\end{equation}
where various positive constants have been absorbed into a new
definition of $\alpha$ and $\beta$.  An instance is \textit{enstrophy
  equipartition}: $\alpha =0$ (all the modes have the same enstrophy),
for which the energy spectrum is $E(k) \propto k^{D-3}$. Following
LPP, if we now set $D=4/3$, we obtain a $k^{-5/3}$ spectrum, the
spectrum predicted by Kolmogorov's 1941 scaling theory and extended by
Kraichnan to the inverse energy cascade of 2D turbulence. Note that
such Gibbs states are only conditionally Gaussian, for a given
disorder.  Otherwise, they are highly intermittent, since a given
high-$k$ mode will be active only in a small fraction of the disorder
realizations. We also note that similar phenomena have been observed
in shell models \cite{aurell-gilbert-etal}.

The form \eqref{Ddimabseq} of the D-dimensional absolute equilibria
also allows for the kind of Bose condensation in the gravest modes
(here, those with unit wavenumber) found by Kraichnan for 2D
turbulence. For this the ``inverse temperature'' $\alpha$ must be
taken negative, close to its minimum realizable value $-\beta$.  The
arguments used by Kraichnan to predict an inverse Kolmogorov
$k^{-5/3}$ energy cascade for high-Reynolds number 2D turbulence with
forcing near an intermediate wavenumber $k_{\rm inj}$ carry over to the
decimated case with $D<2$. In particular the conservation of enstrophy
blocks energy transfer to high wavenumbers. This in itself does not
imply that the energy will cascade to wavenumbers smaller than $k_{\rm
  inj}$, producing a $k$-independent energy flux: it might also linger
around and accumulate near $k_{\rm inj}$.

It is now our purpose to show that for $4/3<D\le 2$, when the energy
spectrum is prescribed to be $E(k) =k^{-5/3}$ over the inertial range,
there is a \textit{negative energy flux} $\Pi_E$, vanishing linearly
with $D-4/3$ near the critical dimension $D=4/3$. For this we shall
assume that a key feature of the two-dimensional energy cascade
carries over to lower dimensions, namely the existence of scaling
solutions with local (in Fourier space) dynamics, so that the energy
transfer is dominated by triads of wavenumbers with comparable
magnitudes. Let us now decompose the energy inertial range into bands
of fixed relative width, say one octave, delimited by the wavenumbers
$2^0$, $2^1$, $2^2$, etc. Because of locality there is much intraband
dynamics but, of course, interband interactions are needed to obtain
an energy flux. Pure intraband dynamics (with no forcing and
dissipation) would lead to thermalization. For dimensional reasons,
thermalization and interband transfer have the same time scale, namely
the eddy turnover time $k^{-3/2} E ^{-1/2}(k)$.

\begin{figure}
\begin{center}
\includegraphics[width=7.cm]{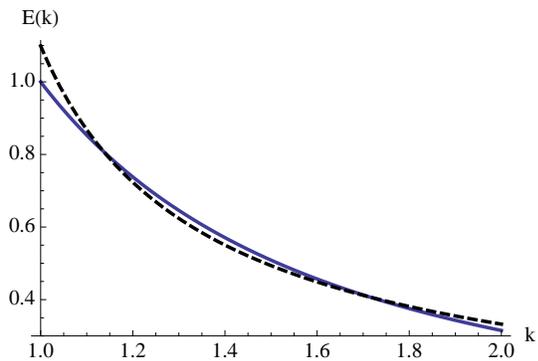}
\end{center}
\caption{(Color online) The $k^{-5/3}$ spectrum (continuous) and the 
associated 2D absolute
  equilibrium with the same energy and enstrophy in the first octave (dashed).}
\label{f:tenandfive}
\end{figure}

To get a handle on the combined intraband and interband dynamics we
perform a thermodynamic thought experiment in which we artificially separate them in
time. In the first phase, starting from a $k^{-5/3}$ spectrum we prevent
the various bands from interacting by introducing (impenetrable)
interband barriers at their edges.  In each band, the modes will then
thermalize and achieve a Gibbs state with a spectrum \eqref{Ddimabseq}
in which $\alpha$ and $\beta$ are 
determined by the constraints that the total band energy and enstrophy 
be the same as for the ${-5/3}$ spectrum. For example, in the first
band this gives the constraints ($n=0$ for the energy and $n=2$ for
the enstrophy)
\begin{equation}
\int_1^2 dk\, k^n\left[k^{D-1}/\left(\alpha +\beta k^2\right)-k^{-5/3}\right] =0.
\label{alphabetaequations}
\end{equation}
In general this constitutes a system of transcendental equations for
the parameters $\alpha$ and  $\beta$ that can only be solved
numerically.  This is illustrated in
Fig.~\ref{f:tenandfive} for the 2D case. 
We observe that in 2D the absolute equilibrium spectrum and the ${-5/3}$
spectrum are very close to each other. Specifically, in 2D the absolute
equilibrium spectrum exceeds the ${-5/3}$ spectrum by about 10\% at
any lower band edge and by about 5\% at any upper band edge. Of
course, as we approach the critical dimension $D=4/3$ the discrepancy
goes to zero and can easily be calculated pertubatively in $D-4/3$. In
the second phase of our thought experiment, we remove just one of the
barriers between two adjacent bands, say, the barrier at $2^1$. A new
thermalization leads then to an absolute equilibrium in the band
$[2^0,\,2^2]$, which again, can be easily calculated. In 2D, before the
removal, the energy between $2^0$ and $2^1$ was $0.555$.  After the
new thermalization, this energy is found to have \textit{increased} by
$0.00551$. Thus energy has been transferred from the upper band
$[2^1,\,2^2]$ to the lower band $[2^0,\, 2^1]$. Close to $D=4/3$, we can
again apply elementary perturbation techniques and obtain for the
upper-to-lower-band energy transfer $0.009 (D-4/3)$ to leading
order. Our thermodynamic thought experiment thus suggests that the
energy flux vanishes linearly with $D-4/3$, being negative above the
critical
dimension, which implies an inverse cascade. In the K41
inertial range, the energy spectrum and the energy flux $\Pi_E$ are
related by $E(k) =C_{\rm Kol} |\Pi_E|^{2/3}
k^{-5/3}$, where $C_{\rm Kol}$ is the Kolmogorov constant, we infer that the Kolmogorov constant diverges
as $(D-4/3)^{-2/3}$. 
 A closure calculation of eddy-damped
 quasi-normal Markovian (EDQNM) type 
also predicts a divergence with a  $-2/3$ exponent.


Kraichnan's ideas about the inverse cascade in 2D got growing support 
a few years later from direct numerical
simulations, which eventually achieved the resolution of $32,768^2$
modes \cite{boffetta-musacchio2010}. As to our idea about
the robustness of the inverse cascade and  the growth of the
Kolmogorov constant when lowering the dimension $D$, some support can
be already provided, using a $D$-dimensional decimated variant of spectral
direct numerical simulation: First one generates
an instance of the disorder, that is the list of active and
inactive Fourier modes; then, one applies standard time marching
algorithms and, at each time step, sets to zero all inactive modes.
In addition to the well-known difficulties of simulating 2D turbulence
(see e.g. \cite{boffetta-musacchio2010} and references therein), there
are new difficulties.

A few words about the numerical implementation. We integrate the
decimated Navier--Stokes equation
\eqref{NSD} in vorticity representation. Instead of using as
damping the viscous operator $\Lambda = \nu \Delta$ (where $\Delta \equiv
\nabla ^2$ is the Laplacian), we use
\begin{equation}  
\Lambda \equiv -\nu \Delta ^{+2} - \mu \Delta ^{-2}, \quad \nu>0, \quad \mu>0,
\label{modifdiss}
\end{equation}
whose Fourier symbol is $-\nu k^4- \mu k^{-4}$. In other words, we  use
hyperviscosity to avoid wasting resolution on the enstrophy
cascade and large-scale friction to prevent an accumulation of energy
on the gravest modes and thus allow eventual convergence to a statistical
steady state. The results reported here have a resolution 
of $N=3072$ collocation points in the two coordinates.
Time marching is done by an Adams--Bashforth scheme combined with exponential
time difference (ETD) \cite{cox-matthews} with a time step  between 
$5\times 10^{-5}$ and $10^{-4}$, depending on dimension.
Energy injection at the rate $\varepsilon$ is done 
in a band of width three around $k_{\rm inj} = 319$ by adding to the
time-rate-of
change of the Fourier amplitude of the vorticity a term proportional to the
inverse of its complex conjugate \cite{shiyi-etalJFM2009}.  This allows a ${\B k}$-independent and
time-independent energy injection. As $D$ is decreased the amplitude of this
forcing is increased to keep the total energy injection on \textit{active} 
modes fixed at $\varepsilon = 0.01$.
 The damping parameters
are $\nu = 10^{-11}$ and $\mu = 0.1$. Runs are done concurrently for
different values of $D$ on a high-performance cluster at the Weizmann 
Institute and take typically a
few thousand hours of CPU per run to achieve a statistical  steady state.

Energy spectra are obtained by angular averages over Fourier-space
shells of unit width
\begin{equation}  
E(K) \equiv \frac{1}{2}\sum_{K\le k<K+1} |\hat{\B v}({\B k})|^2, 
\label{defEk}
\end{equation}
where the $\hat{\B v}({\B k})$ are the Fourier coefficients of the
solution of the decimated Navier--Stokes equation \eqref{NSD}. We also
need the energy flux $\Pi_E(K)$ through wavenumber $K$ due to
nonlinear transfer, defined as
\begin{equation}  
  \Pi_E(K) \equiv \sum_{k\le K}\hat{\B v^*}({\B k})\cdot \widehat{{\rm
      NL}}({\B k}), 
\label{defPi}
\end{equation}
where $\widehat{{\rm  NL}}({\B k})$ denotes  the set of Fourier coefficients of the
nonlinear term $P_DB({\B v},{\B v})$ in the decimated Navier--Stokes equation
\eqref{NSD} and the asterisk denotes complex conjugation. $E(k)$
and $\Pi_E(K)$ are mostly insensitive to the disorder realization. 

\begin{figure}
\begin{center}
\includegraphics[width=6.9cm]{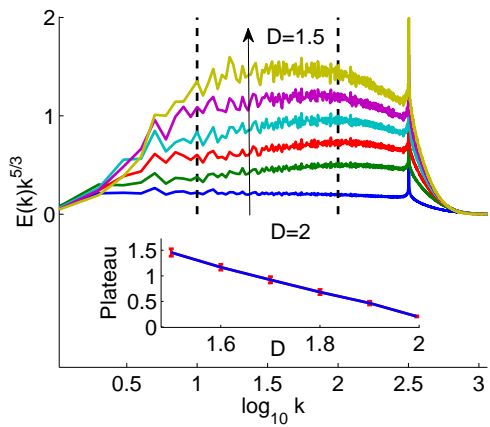}
\end{center}
\caption{(Color online)  Compensated steady-state spectra for
  $D=2.0,\, 1.9,\, 1.8,\, 1.7,\, 1.6,\, 1.5$ from bottom to top with spikes at injection. 
The inset shows the dependence on $D$ of the plateau of the compensated
spectra, as an average over the interval between
vertical dashed lines (with standard deviation error bars).}
\label{f:comp-en-spectra}
\end{figure}
\begin{figure}
\begin{center}
\includegraphics[width=6.9cm]{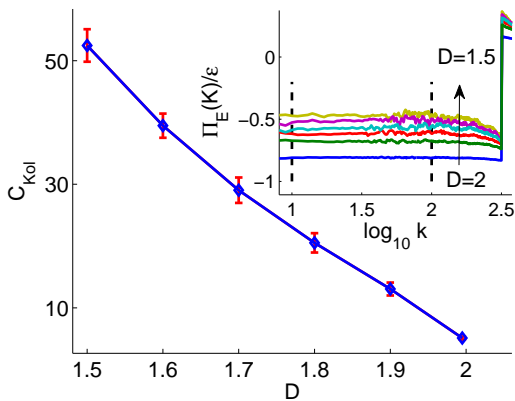}
\end{center}
\caption{(Color online) Dependence of the Kolmogorov constant on
  $D$. The lowest value, at $D=2$, is about 5. The inset shows the
  energy flux 
normalized by the energy injection $\varepsilon$ for the same values of $D$ as
  in Fig.~\ref{f:comp-en-spectra}.}
\label{f:energy-flux}
\end{figure}

Figs.~\ref{f:comp-en-spectra} and \ref{f:energy-flux} (inset) show the steady-state
\textit{compensated} energy spectra $k^{5/3} E(k)$ and the energy fluxes
$\Pi_E(k)$, for various values of $D$ between $2$ and $1.5$, 
respectively. 
Both are quite flat, over a significant range of $k$ values, evidence that $D$-dimensional forced turbulence, Fourier
decimated down from the two-dimensional case, preserves the key
feature of two-dimensional turbulence of having an inverse cascade
that follows the ${-5/3}$ law. Note that the inertial range (the
flat region of the compensated energy spectrum) shrinks as the
dimension $D$ decreases.  The absolute value of the energy flux is
about 80\% of the energy injection $\varepsilon$ for $D=2$, but drops
to less than 50\% for $D=1.5$. Indeed, as the dimension $D$ is
lowered, there are fewer and fewer pairs of \textit{active} modes in
the forcing band, capable through their beating interaction of
draining the energy into the infrared direction; thus the energy
injection will be more and more balanced by direct dissipation near
injection. Preventing this would require a substantial lowering of the
dissipation which in turn requires a substantial increase in the
resolution at the high-$k$ end. Anyway, the fact that the flux
$|\Pi_E|$ becomes substantially lower than injection does not prevent
us from calculating the Kolmogorov constant, given (in terms of plateau
values) by $C_{\rm Kol} = k^{5/3}E(k)/\left(|\Pi_E(k)|^{2/3}\right)$.
Figure~\ref{f:energy-flux} shows the variation of the Kolmogorov
constant with dimension. When lowering the dimension from 2 to 1.5, a
combined effect of a rise in the compensated spectrum and a drop in
flux yields a monotonic growth of about a factor ten in the Kolmogorov
constant and a substantial growth of errors due to fluctuations within the
averaging interval. Probing the conjectured divergence by moving closer to the
critical point $D=4/3$ would require much higher resolution. A
state-of-the-art $16,384^2$ simulation of sufficient length might shed
light.

We finally observe that the fractal Fourier decimation procedure can
be started from any integer dimension and can be applied to a large
class of problems in hydrodynamics and beyond. It could be
interesting, for example, to investigate how it affects the
dissipative anomaly of shock-dominated compressible flow.

We are grateful to E.~Aurell, H.~Chen, H.~Frisch, B.~Khesin, V.~L'vov,
T.~Matsumoto, S.~Musacchio, R.~Pandit and W. Pauls for useful
discussions. U.F. and S.S.R.'s work was supported by ANR ``OTARIE''
BLAN07-2\_183172. A.P. and I.P.'s work was supported by the Minerva
Foundation, Munich, Germany.


\end{document}